\documentclass{ws-procs9x6}

\begin{document}
\input psfig.sty

\def\n16{$^{16}N$}
\def\xo16{$^{16}O$}
\def\xco2{$CO_2$}
\def\c12ag{$^{12}C(\alpha,\gamma)^{16}O$}
\def\go16{$^{16}O(\gamma,\alpha)^{12}C$}
\def\sE1{$S_{E1}(300)$}
\def\xsE2{$S_{E2}(300)$}

\title{OPEN QUESTIONS IN STELLAR HELIUM BURNING STUDIED WITH REAL PHOTONS 
\footnote{Work Supported by USDOE Grant No. DE-FG02-94ER40870.}}

\author{Moshe Gai}

\address{Laboratory for Nuclear Sciences, Department of Physics, U3046, \\
University of Connecticut, 2152 Hillside Rd., Storrs, CT 06269-3046, USA \\ 
E-mail: gai@uconn.edu URL: http://www.phys.uconn.edu}

%%%%%%%%%%%%%%%%%%%%%%%%%%%%%%%%%%%%%%%%%%%%%%%%%%%%%%%%%%%%%%
% You may repeat \author \address as often as necessary      %
%%%%%%%%%%%%%%%%%%%%%%%%%%%%%%%%%%%%%%%%%%%%%%%%%%%%%%%%%%%%%%

\maketitle

\abstracts{
The outcome of helium burning is the formation of the two elements, 
carbon and oxygen.  The ratio of carbon to oxygen at the end of helium 
burning is crucial for understanding the final fate of a progenitor 
star and the nucleosynthesis of heavy elements in Type II supernova,
with oxygen rich star predicted to collapse to a black hole, and a carbon 
rich star to a neutron star. Type Ia supernovae (SNeIa) are used as 
standard candles for measuring cosmological distances with the use of an 
empirical light curve-luminosity stretching factor.  It is essential 
to understand helium burning that yields the carbon/oxygen white 
dwarf and thus the initial stage of SNeIa.  Since the triple alpha-particle 
capture reaction, $^{8}Be(\alpha,\gamma)^{12}C$, the first burning stage in 
helium burning, is well understood, one must extract the cross section of 
the $^{12}C(\alpha,\gamma)^{16}O$ reaction at the Gamow peak (300 keV)
with high accuracy of approximately $10\%$ or better. This goal has not 
been achieved despite repeated strong statements that appeared in the 
literature. Constraint from the beta-delayed alpha-particle emission of 
$^{16}N$ were shown to not sufficiently restrict the p-wave cross section 
factor; e.g. a low value of $S_{E1}(300)$ can not be ruled out. Measurements 
at low energies, are thus mandatory for determining the elusive cross section 
factor for the \c12ag reaction. We are constructing a Time Projection Chamber 
(TPC) for use with high intensity photon beams extracted from the HI$\gamma$S/TUNL 
facility at Duke University to study the \go16 reaction, and thus the direct reaction 
at energies as low as 0.7 MeV. This work is in progress.}

\section{Introduction: Oxygen Formation in Helium 
  Burning and The \c12ag Reaction}

The outcome of helium burning is the formation of the two elements, carbon
and oxygen \cite{Fo84,We93,Ga99}.  The ratio of carbon to oxygen at the end 
of helium burning is crucial for understanding the fate of Type II supernovae 
and the nucleosynthesis of heavy elements. While an oxygen rich star
is predicted to end up as a black hole, a carbon rich star leads to
a neutron star \cite{We93}. At the same time helium burning is also 
very important for understanding Type Ia supernovae (SNeIa) that are now 
being used as a standard candle for cosmological distances \cite{SNIa}.
All thus far luminosity calibration curves and the stretching factor 
are based on empirical observations without fundamental understanding the 
relation between the time characteristics of the light curve and the maximum  
magnitude of Type Ia supernova. Since the first burning stage in helium burning, 
the triple alpha-particle capture reaction ($^{8}Be(\alpha,\gamma)^{12}C$), 
is well understood \cite{Fo84}, one must extract the p-wave [\sE1] and 
d-wave [\xsE2] cross section of the $^{12}C(\alpha,\gamma)^{16}O$  reaction 
at the Gamow peak (300 keV) with high accuracy of approximately $10\%$  or 
better to completely understand stellar helium burning, and better understand 
Type II and Type Ia supernova.

\subsection{Beta-Delayed Alpha-Particle Emission of \n16}

Early hopes for extracting the astrophysical E1 S-factor
[\sE1 $=\sigma _{E1} \times E \times e^{2\pi \eta}$] of the \c12ag reaction
through constraints imposed by new data on the beta-delayed alpha-particle
emission of \n16 \cite{Zh93,Zh93a,Bu93,Az94} were examined in detail over the 
last few years and a few observations were made. The original Yale data 
\cite{Zh93,Zh93a} were improved \cite{Fr96,Fr96a} in a phase II experiment of 
the Yale-UConn group, and were found to be in disagreement with the TRIUMF data 
\cite{Az94} but consistent with the unpublished data of the Seattle group 
\cite{Zh95}, see Fig. 1. In addition, an independent R-matrix analysis
\cite{Ha96} of the world data including the beta decay of $^{16}N$ data   
was found to not rule out a small S-factor solution (10-20 keV-b). It is thus 
doubtful that one can extract the p-wave cross section factor with a reasonable 
accuracy as stated in Ref. \cite{Az94}. The confusion in this 
field mandates a direct measurement of the cross section of the \c12ag reaction 
at low energies. A recent measurement at lower energies \cite{Hammer01}, suggest a 
d-wave cross section factor that is at least twice larger than "the 
accepted value", and the their low energy data point(s) measured with low precision 
can not rule out a small p-wave cross section factor.

\centerline{\psfig{figure=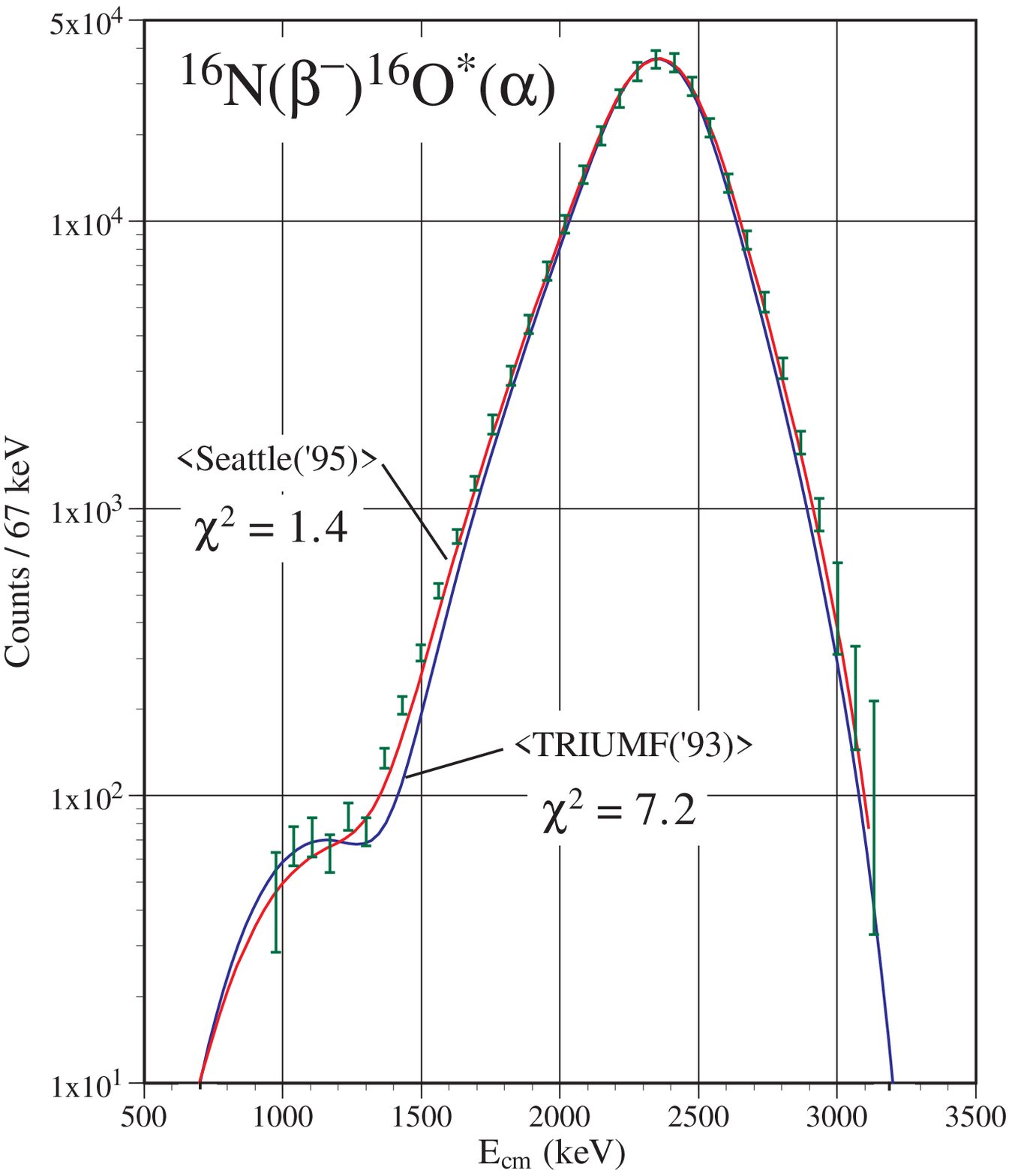,height=4in}}

\hspace{0.5in} \underline{Fig. 1:}  The Yale-UConn data on the beta-delayed 
alpha-particle emission of $^{16}N$ \cite{Fr96,Fr96a} compared to the TRIUMF 
\cite{Az94} and Seattle results \cite{Zh95}. The TRIUMF and Seattle data are 
averaged over the energy resolution of the Yale-UConn experiment and are shown 
by continuous lines. The unpublished Seattle data are listed (by permission) in 
the appendix of Ref. \cite{Fr96a}.

\section{The Proposed \go16 Experiments}

For determining the cross section of the \c12ag at very low energies,
as low as $E_{cm}=700$ KeV, considerably lower than measured till now
\cite{Hammer01}, it is advantageous to have an experimental setup with larger 
(amplified) cross section, high luminosity and low background.  It turns
out that the use of the inverse process, the \go16 reaction may
indeed satisfy all three conditions. The cross section of
\go16 reaction (with polarized photons) at the kinematical region
of interest (photons aprox. 8-8.5 MeV) is larger by a factor of 
approximately 100 than the cross section of the direct \c12ag reaction. 
Note that the linear polarization of the photons yields an extra factor of 
two in the enhancement due to detailed balance. Thus for the lowest thus far 
measured data point at 0.9 MeV with the direct cross
section of aprox.. 60 pb, the photodissociation cross section is 6 nb. It
is evident that with similar luminosities and lower background, see below, 
the photodissociation cross section can be measured to yet lower center of 
mass energies, as low as 0.7 MeV, where the direct \c12ag reaction cross section 
is of the order of 1 pb. A very small contribution (less than 5\%) from cascade 
gamma decay can not be measured in this method, but appears to be negligible  
and below the design goal accuracy of our measurement of $\pm 10\%$.

\centerline{\psfig{figure=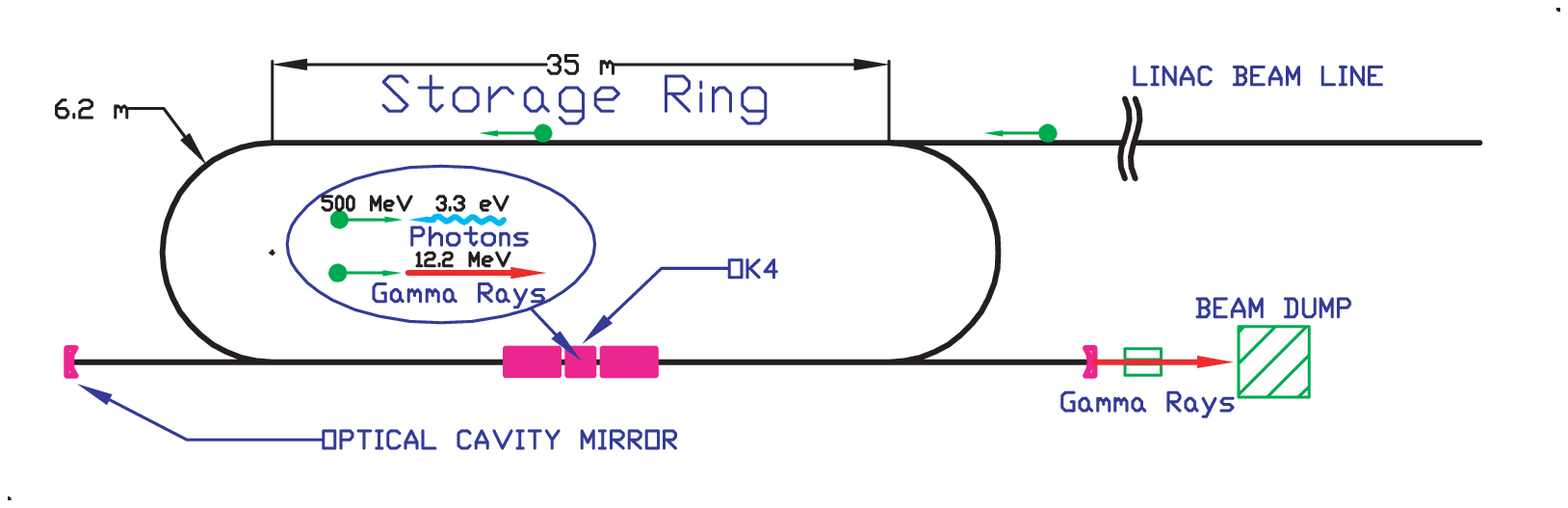,width=6in}}

\hspace{0.5in} \underline{Fig. 2:}  Schematic diagram of the HI$\gamma$S 
     facility \cite{HIGS} for the production of intense MeV gamma beams.

The High Intensity Gamma Source (HI$\gamma$S) \cite{HIGS}, shown in Fig. 2, 
has already achieved many of its design milestones and is rapidly
approaching its design goal for 2-200 MeV gammas. For 9 MeV gammas we 
expect an energy resolution of 0.1\% and intensity of order $10^9$ /sec. 
Currently achieved intensities are of the order of $10^8$ /sec with energy 
resolution of aprox.. 0.5\%. The backscattered photons of the HI$\gamma$S 
facility are collimated (3 mm diameter) and enter the target/detector TPC 
setup as we discuss below. With a Q value of -7.162, our experiment will 
utilize gammas of energies approximately 8 to 10 MeV.  Note 
that the emitted photons are linearly polarized \cite{Lit97} and the 
emitted particles are primarily in a horizontal plane (parallel to 
target room floor) with a $sin^2 \phi$ azimuthal 
angular dependence \cite{PRCRC}, thus simplifying the tracking 
of particles in this experiment. The pulsed photon beam (0.1 ns every 180 nsec
with at most 500 gammas per pulse) 
provides additional trigger for removing background. The image intensified CCD 
camera is triggered by light detected in the PMT, see below, and the time 
projection information from the drift chamber yields the azimuthal angle
of the event of interest. The scattering angle is measured with high accuracy 
using the (8 cm long) alpha tracks and (2 cm long) carbon tracks. Background 
events from contaminants carbon, oxygen and fluorine isotopes, ar 
discriminated using the TPC as a calorimeter with a 2\% energy 
resolution. Time of flight techniques, and flushing
of the CCD between two events will also be used. To reduce noise, the
CCD will be cooled. We note that similar research program with 
high intensity photon beams and a TPC already exists at the 
RCNP at Osaka, Japan \cite{Shima}, proving that tracks from low energy 
light ions can be identified in the TPC with a manageable electron 
background. An $^{16}O(e,e' \alpha)^{12}C$ 
experiment with virtual photons proposed at the MIT-Bates accelerator 
\cite{Genya} is useful to extract only the d-wave astrophysical cross 
section factor and thus it complements our experiment proposed for 
the HI$\gamma$S-TUNL facility \cite{Phys}.

\subsection{Proposed Time Projection Chamber (TPC)}

\centerline{\psfig{figure=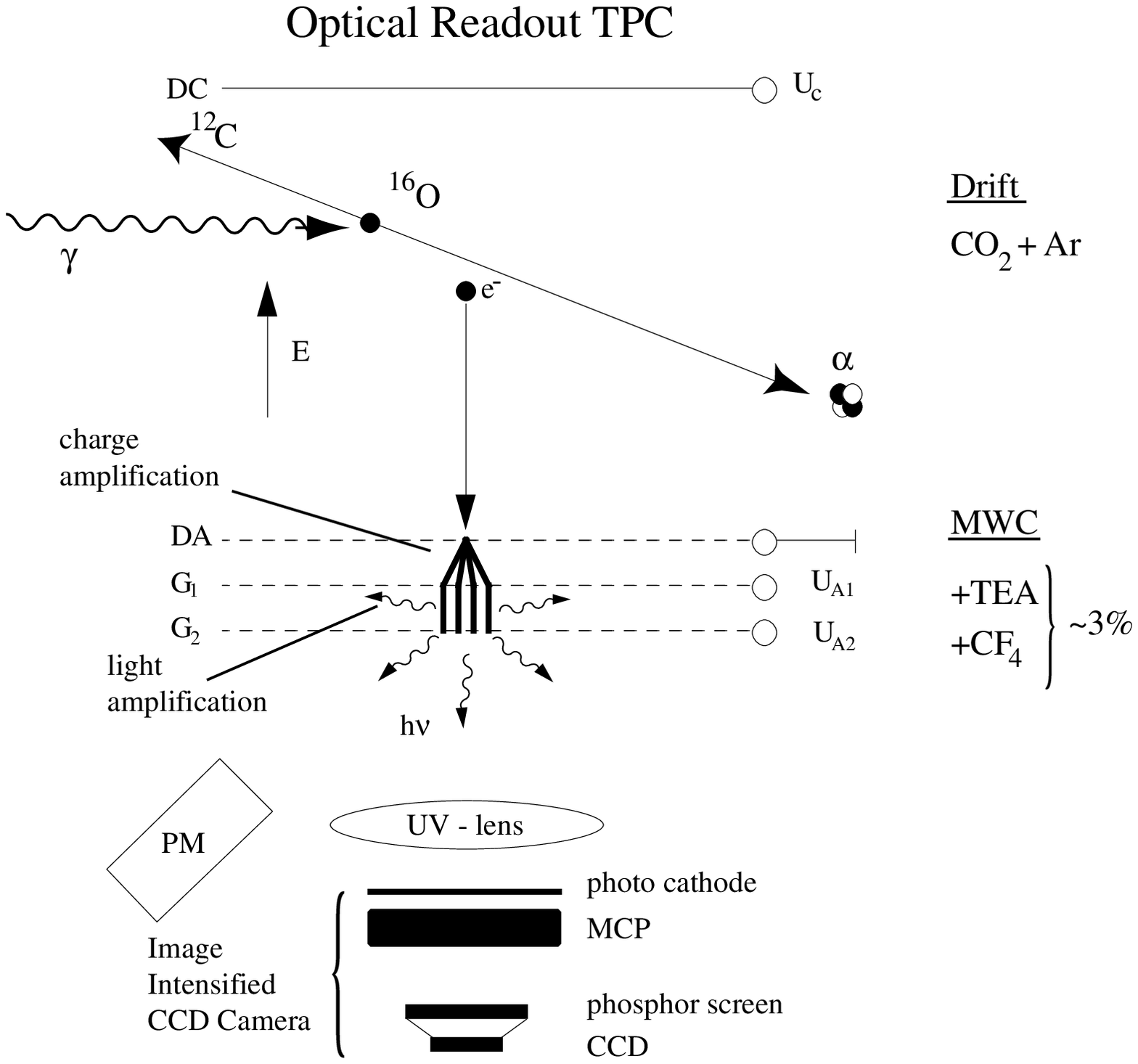,height=3in}}

\hspace{0.5in} \underline{Fig. 3:}  Schematic diagram of the Optical
      Readout TPC \cite{NIMA}.

We are constructing an Optical Readout Time Projection Chamber (TPC), 
similar to the TPC constructed in the Physikalisch Technische Bundesanstalt,
(PTB) in Braunschweig, Germany and the Weizmann Institute, Rehovot, 
Israel \cite{NIMA}, for the detection of alphas and carbon, the byproduct 
of the photodissociation of \xo16.  Since the range of available alphas is 
approximately 8 cm (at 100 mbars) the TPC is 40 cm wide and up to 
one meter long. We first construct a 40 cm long TPC for 
initial use at the HI$\gamma$S beam line at TUNL/Duke. The TPC 
is largely insensitive to single Compton electrons, and the large compton  
electron flux, if a problem, can be blocked using a standard beam 
blocker placed between the drift chamber 
volume and the Multi Wire Proportional Counter of the TPC. The TPC 
allows for tracking of both alphas and carbons 
emitted almost back to back from the beam position 
in time correlation. The very different range of alphas and carbons (aprox..
a factor of 4), and differences in the lateral ionization
density, will aid us in particle identification. The TPC also allow us to 
measure angular distributions with respect to the photon beam  
thus separating the E1 and E2 components of the \c12ag reaction. The 
excellent energy resolution of the TPC (aprox. 2\%) allows us to exclude 
events from the photodissociation of nuclei other than $^{16}O$, including 
isotopes of carbon, oxygen and fluorine, that are present in the gas.  
In Fig. 3, taken from Titt {\em et al.} \cite{NIMA}, we show 
a schematic diagram of the Optical Readout TPC. 

The  photon beam enters the TPC through an 
entrance window in the drift chamber part of 
the TPC and mainly produce background $e^+e^-$ pairs and a smaller 
amount of Compton electrons, as well as the photodissociation 
of various nuclei present in the $CO_2 \ + \ Ar$ gas mixture, including 
$^{16}O$. The charged particle byproducts of the 
photodissociation create delta electrons 
that create secondary electrons that drift 
in the chamber electric field with a total time of the order of 1 $\mu s$ 
per 5 cm. The time projection of the drift electrons allows us to measure the 
inclination angle ($\phi$) of the plane of the byproducts, and the tracks themselves 
allow for measurement of the scattering angle ($\theta$), both with an angular  
resolution better than two degrees. The electrons that 
reach the multi-wire chamber are 
multiplied (by aprox. a factor of $10^5$) and interact with a small (3\%) 
admixture of triethylamine (TEA) \cite{NIMA} or $CF_4$ \cite{NIMA2} 
gas to produce UV or visible photons, respectively.
The light detected in the photomultiplier tube, see Fig. 3, triggers the 
Image Intensifier and CCD camera which takes a picture of the visible tracks.
The picture is downloaded to a PC and analyzed for recognition of the two 
back-to-back alpha-carbon tracks originating from the beam position. The 
background electrons lose aprox. 0.5 KeV/cm in the TPC and are removed by 
an appropriate threshold in the trigger Photo Multiplier Tube (PMT).  
Events from the photodissociation of nuclei 
other than $^{16}O$ are removed by measuring the total energy (Q-value) of the 
event with a resolution of 2\%.

\subsection{Design Goals}

The luminosity of our proposed \go16 experiment 
can be very large. For example, with a 30 cm long fiducial length target with 
30\% $CO_2$ at a pressure of 76 torr (100 mbar) and a photon beam of $2 \times 
10^9$ /sec, we obtain a luminosity of $10^{29}\ sec^{-1}cm^{-2}$, or a day long 
integrated luminosity of 10 nb$^{-1}$.  Thus a measurement of the 
photodissociation of $^{16}O$ with cross section of 1 nb, 
yields 10 count per day, leading to a sensitivity for measuring the direct 
\c12ag reaction with a cross section as low as 10 pb, corresponding to energies 
as low as 700 keV. The construction and tests of the TPC is in progress at the 
University of Connecticut, at the PTB in Braunschweig and the Weizmann Institute. 
A mark I experiment to measure coincidences between $\alpha$-particles and 
$^{12}C$ is in progress at the TUNL/HI$\gamma$S facility.

\section{Acknowledgements}

The author would like to acknowledge discussions with Amos Breskin and the 
work of Steve O. Nelson and Joseph A. Dooley on the design and construction 
of the TPC. This work is in collaboration with Henry Weller of the 
TUNL/(HI$\gamma$S) facility.

\end{document}